\documentclass[aps,prl,superscriptaddress,twocolumn,showpacs]{revtex4-2}
\usepackage{dcolumn} 
\usepackage{graphicx}
\usepackage{bm}      
\usepackage{color}
\usepackage{amsmath,amssymb}
\usepackage{hyperref}
\usepackage{multirow}
\usepackage{epstopdf}
\usepackage{latexsym}
\usepackage{subfigure}
\usepackage{wasysym} 
\usepackage{times}
\usepackage{gensymb}
\usepackage{textcomp}

\begin{document}

\title{Non-magnetic ion site disorder effects on the quantum magnetism of a spin-1/2 equilateral triangular lattice antiferromagnet}

\author{Q. Huang}
\email{qhuang11@vols.utk.edu}
\affiliation{Department of Physics and Astronomy, University of Tennessee, Knoxville, Tennessee 37996, USA}

\author{R. ~Rawl}
\affiliation{Department of Physics and Astronomy, University of Tennessee, Knoxville, Tennessee 37996, USA}

\author{W. W. Xie}
\affiliation {Department of Chemistry and Chemical Biology, Rutgers University, Piscataway, NJ, 08854, USA}

\author{E.~S.~Chou}
\affiliation{National High Magnetic Field Laboratory, Florida State University, Tallahassee, FL 32310, USA}

\author{V. S. Zapf}
\affiliation{National High Magnetic Field Laboratory, Los Alamos National Laboratory, Los Alamos, New Mexico 87545, USA}

\author{X. X. Ding}
\affiliation{National High Magnetic Field Laboratory, Los Alamos National Laboratory, Los Alamos, New Mexico 87545, USA}

\author{C.~Mauws}
\affiliation{Department of Chemistry, University of Manitoba, Winnipeg, Manitoba R3T 2N2, Canada}

\author{C.~R.~Wiebe}
\affiliation{Department of Chemistry, University of Winnipeg, Winnipeg, Manitoba R3B 2E9, Canada}
\affiliation{Department of Chemistry, University of Manitoba, Winnipeg, Manitoba R3T 2N2, Canada}
\affiliation{Department of Physics and Astronomy, McMaster University, Hamilton, Ontario L8S 4M1, Canada}

\author{E. X. Feng}
\affiliation{Neutron Scattering Division, Oak Ridge National Laboratory, Oak Ridge, Tennessee 37831, USA}

\author{H. B. Cao}
\affiliation{Neutron Scattering Division, Oak Ridge National Laboratory, Oak Ridge, Tennessee 37831, USA}

\author{W. Tian}
\affiliation{Neutron Scattering Division, Oak Ridge National Laboratory, Oak Ridge, Tennessee 37831, USA}

\author{J.~Ma}
\affiliation{Laboratory of Artificial Structures and Quantum Control, School of Physics and Astronomy, Shanghai Jiao Tong University, Shanghai 200240, China}
\affiliation{Shenyang National Laboratory for Materials Science, Institute of Metal Research, Chinese Academy of Sciences, 110016 Shenyang, China}

\author{Y. ~Qiu}
\affiliation{NIST Center for Neutron Research, National Institute of Standards and Technology, Gaithersburg, Maryland 20899, USA}

\author{N. ~Butch}
\affiliation{NIST Center for Neutron Research, National Institute of Standards and Technology, Gaithersburg, Maryland 20899, USA}

\author{H.~D.~Zhou}
\affiliation{Department of Physics and Astronomy, University of Tennessee, Knoxville, Tennessee 37996, USA}
\affiliation{National High Magnetic Field Laboratory, Florida State University, Tallahassee, FL 32310, USA}

\date{\today}

\begin{abstract}
With the motivation to study how non-magnetic ion site disorder affects the quantum magnetism of Ba$_3$CoSb$_2$O$_9$, a spin-1/2 equilateral triangular lattice antiferromagnet, we performed DC and AC susceptibility, specific heat, elastic and inelastic neutron scattering measurements on single crystalline samples of Ba$_{2.87}$Sr$_{0.13}$CoSb$_2$O$_9$ with Sr doping on non-magnetic Ba$^{2+}$ ion sites. The results show that Ba$_{2.87}$Sr$_{0.13}$CoSb$_2$O$_9$ exhibits (i) a two-step magnetic transition at 2.7 K and 3.3 K, respectively; (ii) a possible canted 120 degree spin structure at zero field with reduced ordered moment as 1.24 $\mu_B$/Co; (iii) a series of spin state transitions for both $H \parallel ab$-plane and $H \parallel c$-axis. For $H \parallel ab$-plane, the magnetization plateau feature related to the up-up-down phase is significantly suppressed; (iv) an inelastic neutron scattering spectrum with only one gapped mode at zero field, which splits to one gapless and one gapped mode at 9 T. All these features are distinctly different from those observed for the parent compound Ba$_3$CoSb$_2$O$_9$, which demonstrates that the non-magnetic ion site disorder (the Sr doping) plays a complex role on the magnetic properties beyond the conventionally expected randomization of the exchange interactions. We propose the additional effects including the enhancement of quantum spin fluctuations and introduction of a possible spatial anisotropy through the local structural distortions.     
\end{abstract}
\maketitle

\section{I. INTRODUCTION}
While quantum magnets are the focus of modern condensed matter physics studies since they provide unique opportunities not only to explore quantum many-body physics but also for applications of advanced technologies, such as spintronics and quantum computers, more attention has recently been paid to their disorder effects since in reality, materials inevitably have defects and/or random disorder. Compared to the disorder on magnetic ion sites, which usually destroys the magnetic interactions and leads to glassy or paramagnetic state, the disorder on non-magnetic ion sites remains more intriguing and illusive. How such kind of disorder affects the magnetism on the chemically ordered magnetic sublattice is a challenging question with limited knowledge so far.

A good example of non-magnetic ion site disorder is the heavily studied YbMgGaO$_4$ (YMGO). In YMGO, the Yb$^{3+}$ ions with effective spin-1/2 form a geometrically frustrated triangular layer\cite{li2015gapless,li2015rare}, between which is the site mixture of Mg$^{2+}$/Ga$^{3+}$ ions causing the intrinsic disorder on the non-magnetic ion sites. The early studies on YMGO, including the continuum mode on the inelastic neutron scattering (INS) spectrum\cite{paddison2017continuous,shen2016evidence,li2017nearest,shen2018fractionalized}, the $C_{\text{p}}$ $\sim T^{0.7}$ behavior for the specific heat\cite{li2015gapless}, the temperature-independent plateau for the Muon spin relaxation (MuSR) rate\cite{li2016muon}, and the saturated DC susceptibility below 0.1 $\sim$ 0.2 K\cite{li2019rearrangement} all suggest a gapless quantum spin liquid (QSL) state in YMGO. Lately, the reports including no residual $\kappa_0$/$T$ term on the thermal conductivity observed\cite{xu2016absence} and the frequency dependent AC susceptibility peak\cite{ma2018spin}, suggest a glassy ground state in YMGO which could be due to the Mg$^{2+}$/Ga$^{3+}$ site disorder\cite{li2015rare,li2017crystalline}. By including this site disorder effect, theoretically, several scenarios have been proposed for YMGO, such as highly anisotropic nearest-neighbor interactions\cite{luo2017ground}, the random singlet (RS) state\cite{kimchi2018valence,parker2018finite}, randomness-induced QSL state\cite{kawamura2019nature}, the mimicry of a spin liquid\cite{zhu2017disorder,zhu2018topography}, and randomness induced spin-liquid-like phase in J$_1$-J$_2$ model\cite{wu2019randomness}. Among them, RS state and randomness-induced QSL have also been applied to other quantum magnets with disorder\cite{kimchi2018scaling,volkov2020random,kundu2020signatures,song2021evidences,syzranov2021hidden,liu2018random,hong2021extreme,baek2020observation,do2020randomly,huang2021heat,hu2021freezing,ren2020characterizing,gomez2021absence,shimokawa2015static,kawamura2014quantum,watanabe2014quantum,uematsu2017randomness,uematsu2018randomness,uematsu2019randomness}. Meanwhile, a most recent study on high quality YMGO single crystals reveals a residual $\kappa_0$/$T$ term and series of quantum spin state transitions at the zero temperature limit, which suggest the survival of itinerant excitations and quantum spin state transitions in YMGO\cite{10.1038/s41467-021-25247-6}.
 
The controversial magnetic ground state of YMGO, again, is a perfect reflection of the complex role that the disorder on the non-magnetic ion sites plays in quantum magnets. To better understand this role, more studies on quantum magnets with such kind of disorder is highly desirable. While searching for ideal quantum magnets to incorporate non-magnetic ion site disorder, Ba$_3$CoSb$_2$O$_9$(BCSO) caught our attention. As a rare example of equilateral triangular lattice antiferromagnet (TLAF) with effective spin-1/2 (for Co$^{2+}$ ions) and easy plane anisotropy, its exotic quantum magnetism have been well documented. First, the theoretical studies have proposed that in a spin-1/2 TLAF, the quantum spin fluctuations stabilize a novel up-up-down (UUD) phase while approaching zero temperature with the applied field parallel to either easy plane or easy axis\cite{miyashita1986magnetic,chubukov1991quantum}, which exhibits itself as a magnetization plateau within a certain magnetic field regime and with one-third of the saturation moment (1/3$M_{\text{s}}$). Experimentally, such a UUD phase has been reported for BCSO\cite{shirata2012experimental, susuki2013magnetization, sera2016s,zhou2012successive}. The detailed studies further reveal series of quantum spin state transitions (QSSTs) in BCSO. With increasing field along the $ab$ plane, its 120 degree spin structure at zero field is followed by a canted 120 degree spin structure, the UUD phase, a coplanar phase (the V phase), and another coplanar phase (the V' phase) before entering the fully polarized state \cite{gekht1997JETP,chen2013groundstates, starykh2014nearsaturation,yamamoto2014quantum,koutroulakis2015quantum, yamamoto2015microscopic,liu2019microscopic}. While for  $H \parallel c$-axis, the 120 degree spin structure will be followed by an umbrella spin structure, and the V  phase. Second, several abnormal features about its spin dynamics have been experimentally studied by the INS experiments \cite{kamiya2018nature,ito2017structure,verresen2019avoided,ma2016static,macdougal2020avoided} and theoretically investigated\cite{mourigal2013dynamical,ghioldi2015magnons,maksimov2016field}, including the quantum renormalization of excitation energies, a rotonlike minimum at the M point, and the intense continuum mode extending to a high energy.

Another fact is that the triple perovskite structure for BCSO is very flexible on its chemical composition. For example, one can replace Ba$^{2+}$ ions with Sr$^{2+}$ or Ca$^{2+}$ ions\cite{lu2018lattice,primo2001synthesis}, or replace Sb$^{5+}$ ions with Nb$^{5+}$ or Ta$^{5+}$ ions\cite{lee2014series,lee2017magnetic,yokota2014magnetic} while keeping the Co-triangular lattice. In fact, Ba$_3$CoNb$_2$O$_9$\cite{lee2014series} exhibits a two-step ordering process at 1.36 K and 1.10 K with easy axis anisotropy and also exhibits the QSSTs. Unfortunately, so far, no single crystalline form of Ba$_3$CoNb$_2$O$_9$ has been prepared, which limits its studies.

These two advantages, exotic quantum magnetism and flexible chemical composition, make BCSO an ideal system to incorporate the non-magnetic ion site disorder and thereafter study its effects on quantum magnetism. In this paper, we successfully grew single crystals of Ba$_{2.87}$Sr$_{0.13}$CoSb$_2$O$_9$ to introduce the disorder (the mixture of Ba/Sr ions) on the non-magnetic Ba$^{2+}$ ion sites and studied its magnetic properties by performing  DC and AC susceptibility, specific heat, elastic and inelastic neutron scattering measurements. By comparing the obtained results to those of the parent compound BCSO, it is learned that such disorder on the non-magnetic ion sites significantly tune the magnetic ground state, phase diagram, and spin dynamics of this spin-1/2 equilateral TLAF.

\section{II. EXPERIMENTAL}
Single crystalline Sr-doped Ba$_3$CoSb$_2$O$_9$ (Sr-BCSO) was synthesized by the floating-zone method. The feed and seed rods for the crystal growth were made by solid state reaction. The appropriate amount of BaCO$_{3}$, SrCO$_3$, CoCO$_3$ and Sb$_2$O$_3$ were ground together, pressed into a 6 mm-diameter 10 cm-long rod under 400 atm hydrostatic pressure, and annealed in air at 1300 $^{\circ}$C for 20 hours. The single crystal growth was carried out in a 10 \%O$_{2}$/90 \%Ar forming gas with a 4 atmosphere gas pressure in an IR-heated image furnace equipped with two halogen lamps. The crystal growth rate is 30 mm/hour. The Laue back diffraction technique is used to align the single crystal sample in specific orientation for measurements. The single crystal x-ray diffraction measurement was performed in Bruker D8 Quest Eco diffractometer with Mo radiation ($\lambda$K$_\alpha$ = 0.71073 $\, $\AA $\,$), which is equipped with Photon II detector. The crystal structure, especially the atomic mixture and disorders, were refined using SHELXTL package with full matrix least-squares on $F^2$ model\cite{sheldrick2015crystal}.

The DC magnetic susceptibility measurements were performed using a vibrating sample magnetometer option in a commercial physical property measurement system (PPMS, Quantum Design) . The specific heat data was also obtained using a PPMS. The magnetization up to 35 T was measured with a vibrating sample magnometer (VSM) at the National High Magnetic Field Laboratory (NHMFL), Tallahassee. Another magnetization measurement up to 40 T using compensated induction coil with sample-in/sample-out background subtraction was performed at the NHMFL, Los Alamos. The single crystal neutron diffraction measurements were performed at the High Flux Isotope Reactor (HFIR) of the Oak Ridge National Laboratory (ORNL). The zero field measurements were carried out at 300 mK with the wavelength of $\lambda$ = 1.5424 $\, $\AA $\,$ in the Oxford cryogen-free HelioxVT refrigerator at the single crystal diffractometer DEMAND\cite{cao2019demand}, HB3A and at 1.5 K with the wavelength of $\lambda$ = 4.2636 $\, $\AA $\,$ in the orange cryostat on the Cold Neutron Triple-Axis Spectrometer, CTAX. The measurements under magnetic field were conducted at 1.5 K with the wavelength of $\lambda$ = 2.3779 $\, $\AA $\,$ in a 8T Vertical Asymmetric Field Cryomagnet on the Fixed-Incident-Energy Triple-Axis Spectrometer, HB1A. The diffraction patterns were analyzed by the Rietveld refinement program FullProf. The magnetic structure compatible with the lattice symmetry was obtained by SARAH software. The INS measurements under zero field were performed at 300 mK on the Disk Chopper Spectrometer (DCS) with the He-3 Dipper insert in the NIST Center for Neutron Research (NCNR). The INS measurements under applied magnetic field were conducted at 1.6 K in an 11T magnet on the Multi Axis Crystal Spectrometer (MACS) in NCNR\cite{rodriguez2008macs}.

\section{III. RESULTS}

\subsection{A. Crystal and magnetic structure}

	\begin{figure}[tp]
	\centering
     {
		\includegraphics[width=3.4in]{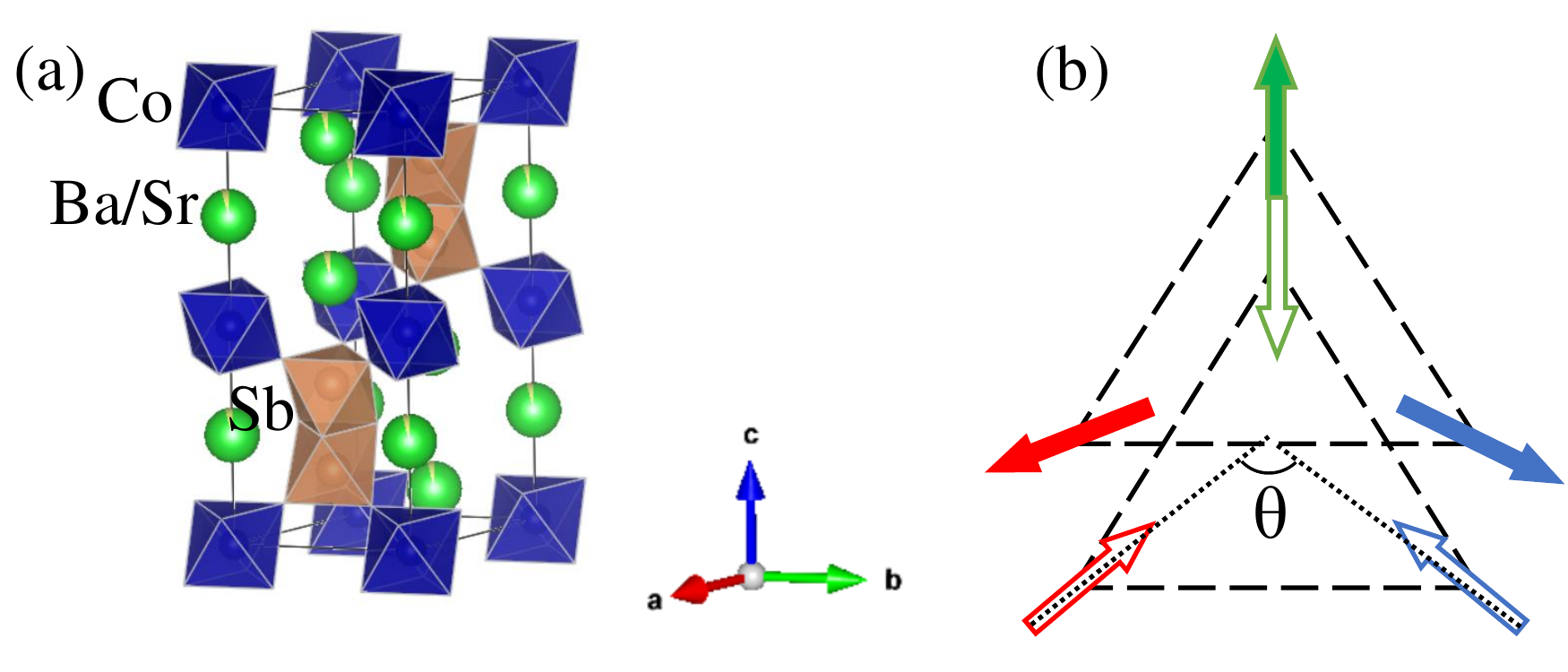}  
		\caption{\label{Fig:1} (color shown online) (a) An illustration of the crystallographic structure of Sr-BCSO. (b) An illustration of the canted 120-degree spin structure.}
	}
\end{figure}

The refinement of the single crystal x-ray data shows that the composition for the as-grown crystal is Ba$_{2.87}$Sr$_{0.13}$CoSb$_2$O$_9$. It has a hexagonal structure with the space group P6$_3$/mmc and lattice parameters $a$ =5.847(3) $\, $\AA $\,$ and $c$ = 14.507(6) $\, $\AA $\,$. The detailed refinement and crystallographic information is listed in table I and II.The results show $\sim4\%$ Sr mixed on Ba site. Moreover, no atomic vacancies were observed on Co, Sb and O sites. As shown in Fig. 1(a), in this structure, each Co$^{2+}$ ion is surrounded by six O$^{2-}$ ions, forming a CoO$_6$ octahedron. The layers of the triangular lattice of CoO$_6$ octahedra in the $ab$ plane are separated by two face-shared SbO$_6$ octahedra. The intralayer and interlayer distances between the Co$^{2+}$ ions are 5.847(3) $\, $\AA $\,$ and 7.253(5) $\, $\AA $\,$, respectively, which lead to a quasi-2D triangular lattice system. 

\begin{figure}[tp]
	\centering
     {
		\includegraphics[width=3.0 in]{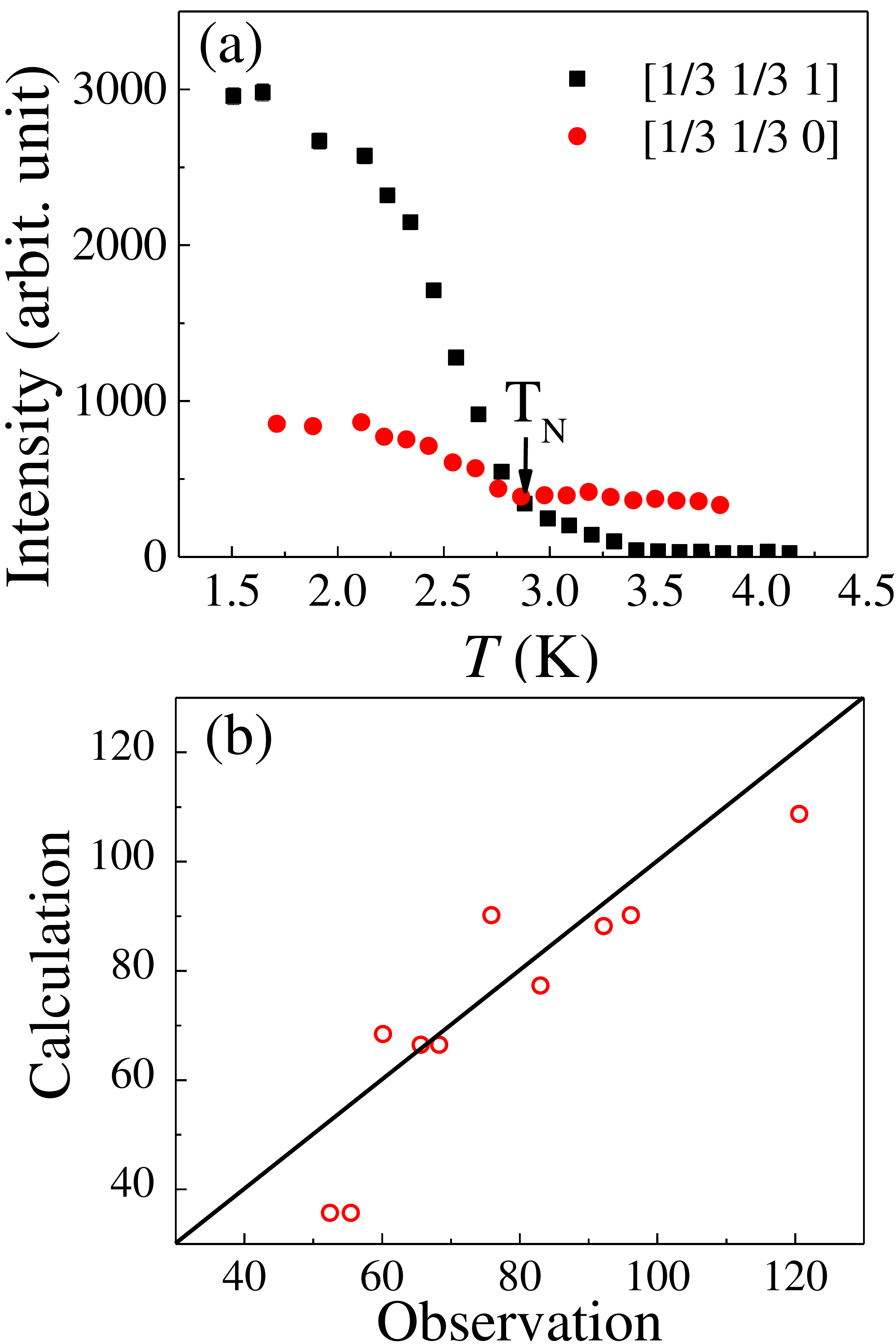}  
		\caption{\label{Fig:2}  (color shown online) (a) The temperature dependence of the intensity for a magnetic peak [1/3 1/3 1] and [1/3 1/3 0]. The intensity of [1/3 1/3 0] is timed by a factor 20 for the visibility. (b) The refinement result of the single crystal neutron diffraction data with a canted 120-degree spin structure.}
	}
\end{figure}

The elastic neutron scattering measurements under zero and applied magnetic fields were performed to probe the magnetic structure of Sr-BCSO. As shown in Fig. 2(a), the temperature dependence of the intensity for a magnetic peak [1/3 1/3 1] measured at zero field shows a rapid increase below $T_{\text{N1}}$ = 2.8 K, which represents a magnetic ordering. Moreover, a sets of magnetic peaks indexed by propagation vector [1/3 1/3 1] were collected in a single crystal neutron scattering measurement. As shown in Fig. 2(b), the data can be fitted by a 120 degree spin structure in the $ab$ plane. The refined magnetic moment is 1.24(7) $\mu_{\text{B}}$/Co. Meanwhile, a weak intensity at zero field for the magnetic peak [1/3 1/3 0] below $T_{\text{N1}}$ was also observed, as shown in Fig. 2(a). The existence of vector [1/3 1/3 0] suggests that spins in the $ab$ plane should have a weak ferromagnetic moment. One possible scenario is the canted 120 degree spin structure, in which the sum of the magnetic moments of the three sublattice spins is nonzero but lead to a resultant magnetic moment, as shown in Fig. 1(b). Unfortunately, the exact spin structure can't be solved due to the limited data resolution we obtained. Therefore, while the data suggests a canted spin structure as the ground state for Sr-BCSO, we still use the 120 degree spin structure to obtain the ordered moment. 

\begin{figure}[tp]
	\centering
     {
		\includegraphics[width=3.0 in]{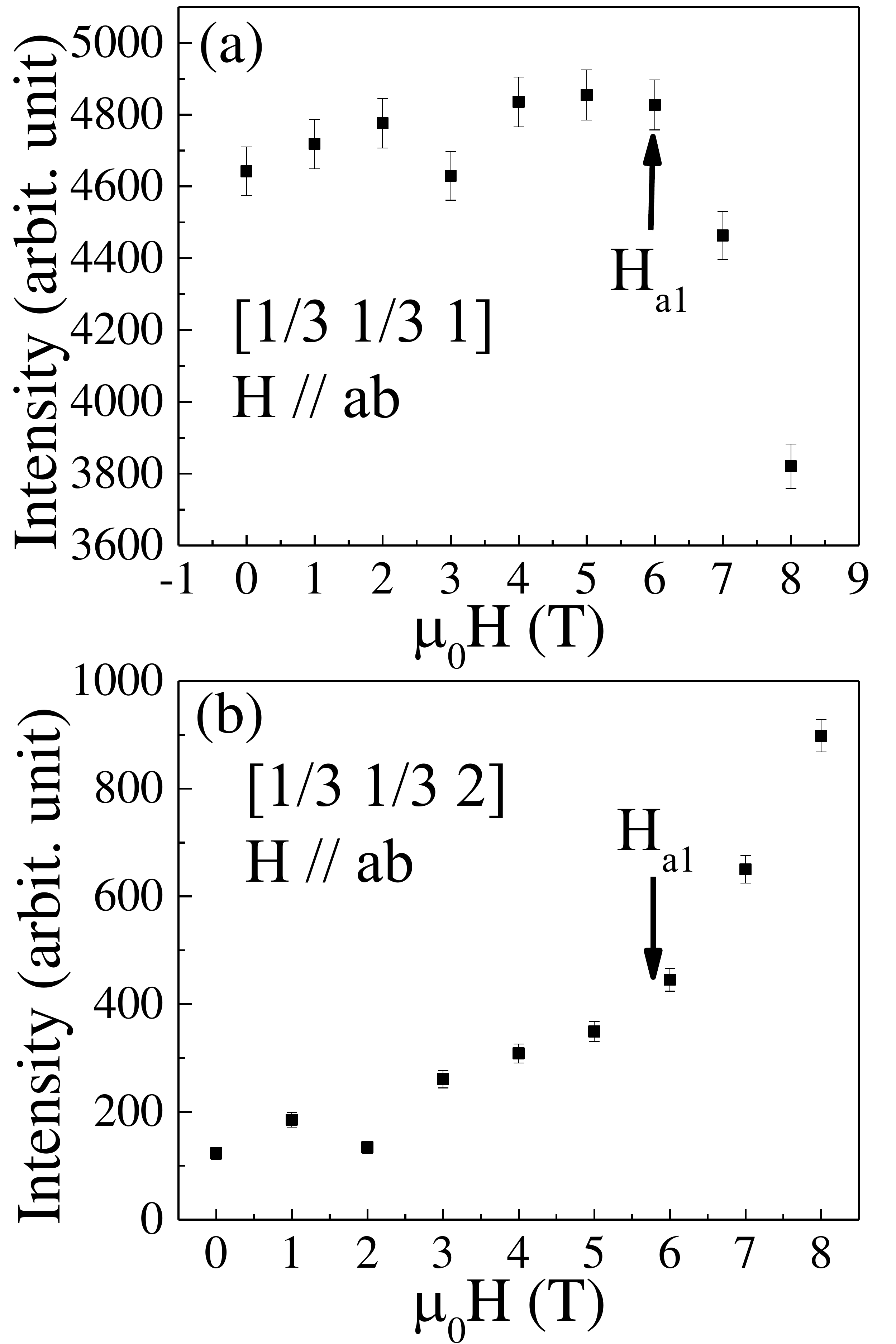}  
		\caption{\label{Fig:3} The field dependence of the intensity for magnetic peak [1/3 1/3 1] (a) and [1/3 1/3 2] (b) measured at 1.5 K.}
	}
\end{figure}

The field dependence of the intensity for the [1/3 1/3 1] and [1/3 1/3 2] peaks measured at 1.5 K are shown in Fig. 3(a) and (b), respectively. For the [1/3 1/3 1] peak, its intensity is basically a constant from 0 T to 6 T, and then shows a drastic drop around $\mu_0H_{\text{a1}}$ = 6 T. For the [1/3 1/3 2] peak, its intensity increases with increasing field but with a slope change around 6 T. These behaviors suggest there is a spin state transition around $H_{\text{a1}}$.

\begin{table}[h]
	\caption{Data of crystallographic refinement for Sr-BCSO at 296 (2)K} 
	\centering 
	\begin{tabular}{c c} 
		\hline\hline 
	    Crystal system & Hexagonal\\ 
		F.W. (g/mol) & 851.74 \\
		Space group;Z & P6$_3$/mmc;2 \\
		a($\, $\AA $\,$) & 5.847 (3) \\ 
		c($\, $\AA $\,$) & 14.507(6) \\
		V($\, $\AA $\,^3$) & 429.5(4) \\
		Extinction Coefficient & 0.00044 (7) \\
		$\theta$ range (deg) & 2.808-34.487 \\
		No. reflections; R$_int$ & 8552; 0.0254 \\
		No. independent reflections) & 394 \\
		No. parameters) & 24 \\
	    R$_1$: $\omega$R$_2$ ($I<2\delta(I)$) & 0.0104; 0.0190 \\
		Goodness of fit & 1.199 \\
		Diffraction peak and hole ($e^-/\, $\AA $\,^3$) & 0.804; -0.800 \\
		\hline 
	\end{tabular}
	\label{table:nuclear} 
\end{table}

\begin{table}[h]
	\caption{Atomic coordinates and equivalent isotropic displacement parameters of Sr-BCSO system.(U$_(eq)$ is defined as one-third of the trace of the orthogonalized U$_(ij)$ tensor ($\, $\AA $\,^2$))} 
	\centering 
	\setlength{\tabcolsep}{0.2mm}
	\begin{tabular}{c c c c c c c} 
		\hline\hline 
		Atom & site & Occ. & x & y & z & U$_(eq)$ \\ [0.5ex] 
		\hline 
		Ba/Sr1 & 2b & 0.952(5)/0.048 & 0 & 0 & 0.25 & 0.009(1)\\ 
		Ba/Sr2 & 4f & 0.955(4)/0.045 & 1/3 & 2/3 & 0.0101(1) \\
		Sb & 4f & 1 & 1/3 & 2/3 & 0.65183(2) & 0.0066(1) \\
		Co & 2a & 1 & 0 & 0 & 0 & 0.0057 (1) \\
		O1 & 6h & 1 & 0.5189(2) & 0.0378(2) & 0.25 & 0.0094(4) \\ 
		O2 & 12k & 1 & 0.8299(2) & 0.6597(3) & 0.0848(1) & 0.0150(3)\\
		\hline 
	\end{tabular}
	\label{table:nuclear} 
\end{table}

\subsection{B. DC and AC magnetic susceptibility}
\begin{figure*}[tp]
	\linespread{1}
	\par
	\begin{center}
		\includegraphics[width= 6.5 in]{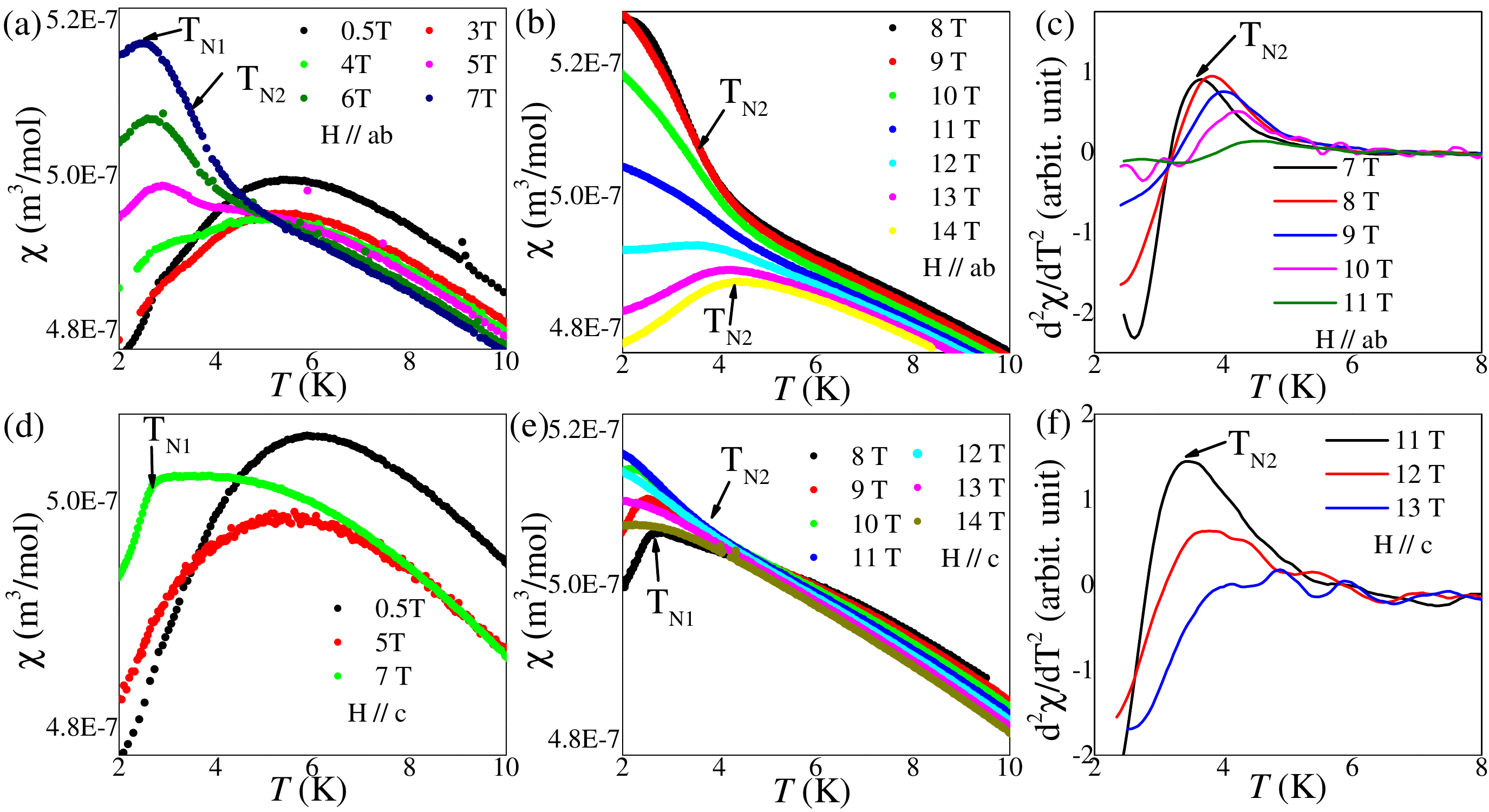}
	\end{center}
	\par
	\caption{\label{Fig:4} (color shown online)  The temperature dependence of the DC susceptibility under different magnetic fields with $H \parallel ab$-plane (a, b) and $H \parallel c$-axis (d, e) for Sr-BCSO. The second-order derivative of the $\chi$-$T$ curves under high magnetic fields for $H \parallel ab$-plane (c) and $H \parallel c$-axis (f).
}
\end{figure*}

The temperature dependence of DC magnetic susceptibility, $\chi(T)$, under magnetic fields is show in Fig. 4. With increasing field $H \parallel ab$-plane, $\chi(T)$ (Fig. 4(a-c)) shows several features including (i) a broad peak appears around 6 K with $\mu_0H$ $<$ 5 T; (ii) another peak appears around 3 K at $\mu_0H$ = 5 T, which is labeled as $T_{\text{N1}}$ and shifts to lower temperatures with increasing field; (iii) at $\mu_0H$ = 7 T, besides the peak at $T_{\text{N1}}$, the data shows a rapid increase below 5 K. Here we use the peak position of its second-order derivative (Fig. 4(c)) to represent this feature, which is labeled as $T_{\text{N2}}$. With increasing field, $T_{\text{N2}}$ shifts to higher temperatures between 7 T $\leq$ $\mu_0H$ $\leq$ 11 T; (iv) with $\mu_0H$ $\geq$ 12 T, the feature at $T_{\text{N2}}$ involves to be a peak again (Fig. 4(b)). For $H \parallel c$-axis, the peak at $T_{\text{N1}}$ is not as strong as that for $H \parallel ab$-plane and appears when $\mu_0H$ $\geq$ 7 T (Fig. 4(d)). The feature at $T_{\text{N2}}$ also appears when $\mu_0H$ $>$ 11 T (Fig. 4(e, f)). Compared to the $H \parallel ab$-plane case, for the $H \parallel c$-axis, both features appear at a higher magnetic field.

\begin{figure}[tp]
	\linespread{1}
	\par
	\begin{center}
		\includegraphics[width= 3.0 in]{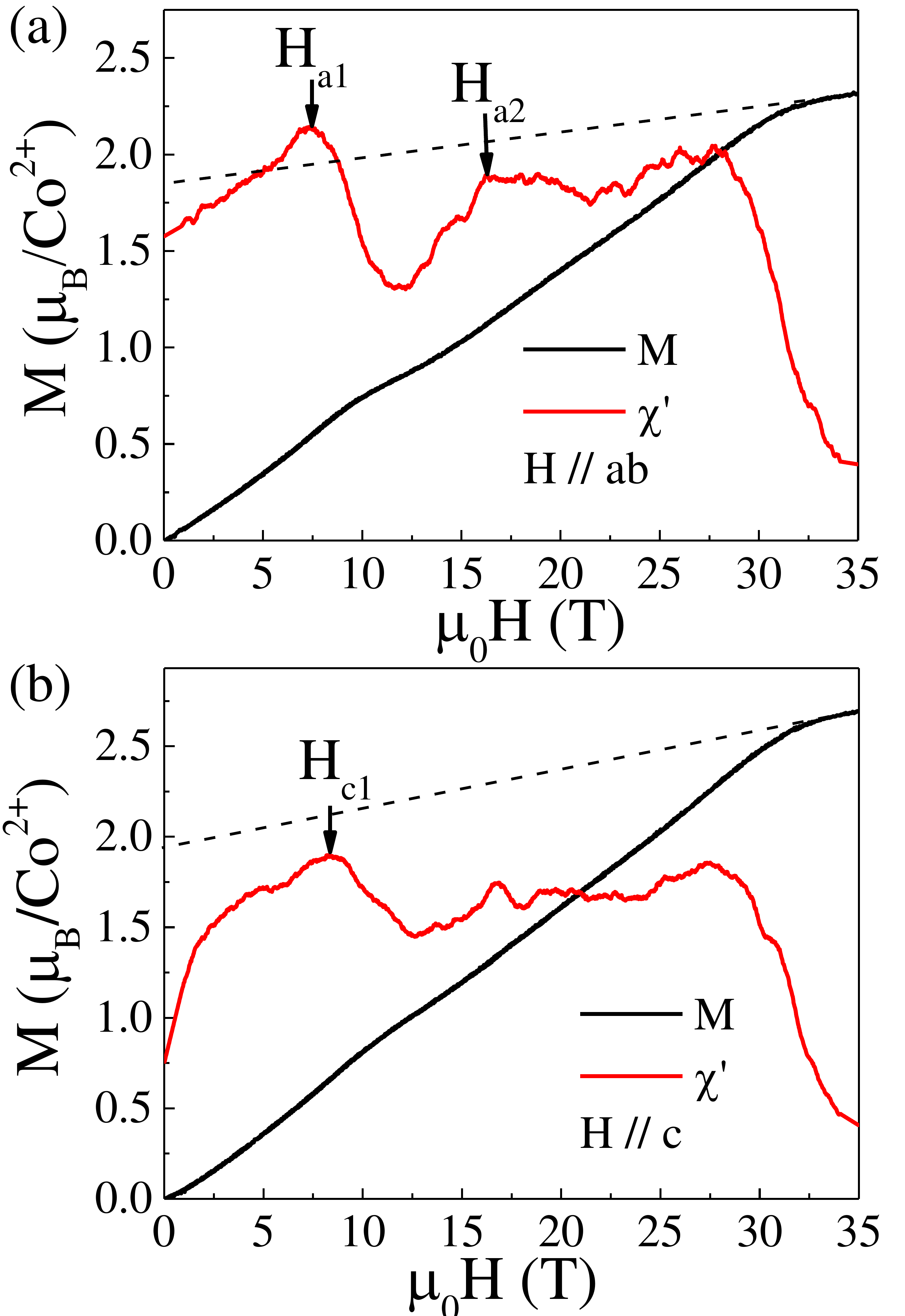}
	\end{center}
	\par
	\caption{\label{Fig:5} (color shown online) The dc magnetization (black line) and its derivative (red line) measured at 1.5 K with $H \parallel ab$-plane (a) and $H \parallel c$-axis (b). The dashed line represents the Van Vleck paramagnetic background. 
}
\end{figure}

As shown in Fig. 5, the DC magnetization, $M(H)$, measured at 1.5 K by using a VSM shows a weak plateau feature for $H \parallel ab$-plane while only a slope change for $H \parallel c$-axis. Accordingly, the derivative d$M$/d$H$ for $H \parallel ab$-plane shows a peak-valley-peak feature with two critical fields $\mu_0H_{\text{a1}}$ = 7.5 T and $\mu_0H_{\text{a2}}$ = 16.4 T. Meanwhile, the derivative for $H \parallel c$-axis only shows one peak at $\mu_0H_{\text{c1}}$ = 8.7 T. The saturation field is around 33 T for both directions. The saturation moments are 1.8 $\mu_{B}$/Co and 1.9 $\mu_{B}$/Co for $H \parallel ab$-plane and $H \parallel c$-axis, respectively. For $H \parallel ab$-plane, the DC magnetization was also measured at 0.7 K by using a compensated induction coil with sample-in/sample-out background subtraction (not shown here), which shows similar behavior as the 1.5 K data.

\begin{figure}[tp]
	\linespread{1}
	\par
	\begin{center}
		\includegraphics[width= 3.0 in]{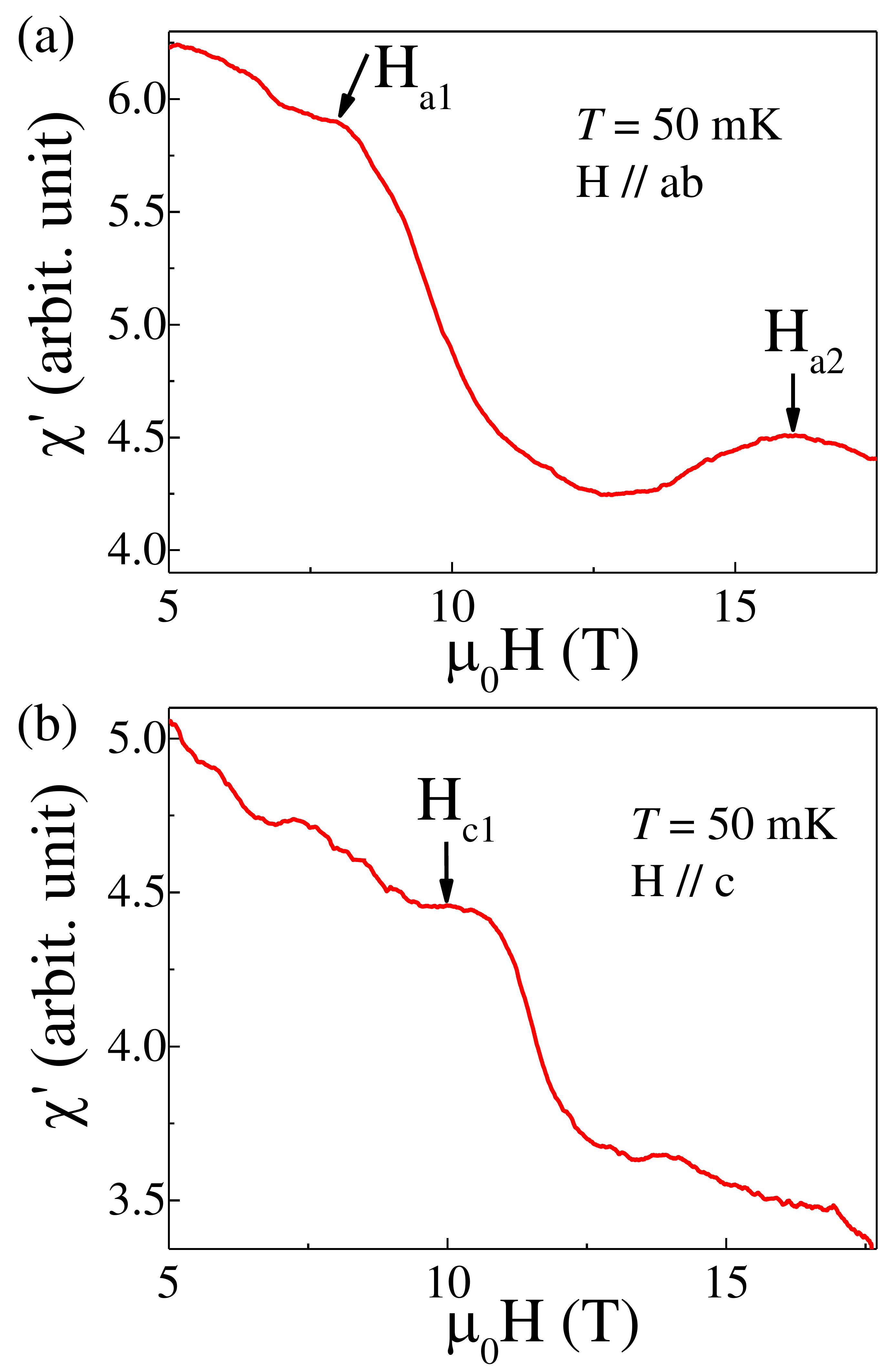}
	\end{center}
	\par
	\caption{\label{Fig:6} (color shown online) The AC susceptibility measured at 50 mK  with  $H \parallel ab$-plane (a) and $c$-axis (b).
}
\end{figure}

Since the field dependent AC magnetic susceptibility, $\chi'(H)$, is basically equivalent to d$M$/d$H$, we measured $\chi'(H)$ at 50 mK to double check the critical fields. As shown in Fig. 6, two critical fields at $\mu_0H_{\text{a1}}$ = 8.1 T and $\mu_0H_{\text{a2}}$ = 16.0 T for $H \parallel ab$-plane and one at $\mu_0H_{\text{c1}}$ = 9.8 T for $H \parallel c$-axis were observed, which are consistent with the DC $M(H)$ results. Since the $\chi'(H)$ signal is on top of the coil background, it appears less pronounced than d$M$/d$H$ in Fig. 5.

\subsection{C. Specific heat}

\begin{figure}[tp]
	\linespread{1}
	\par
	\begin{center}
		\includegraphics[width= 3.0 in]{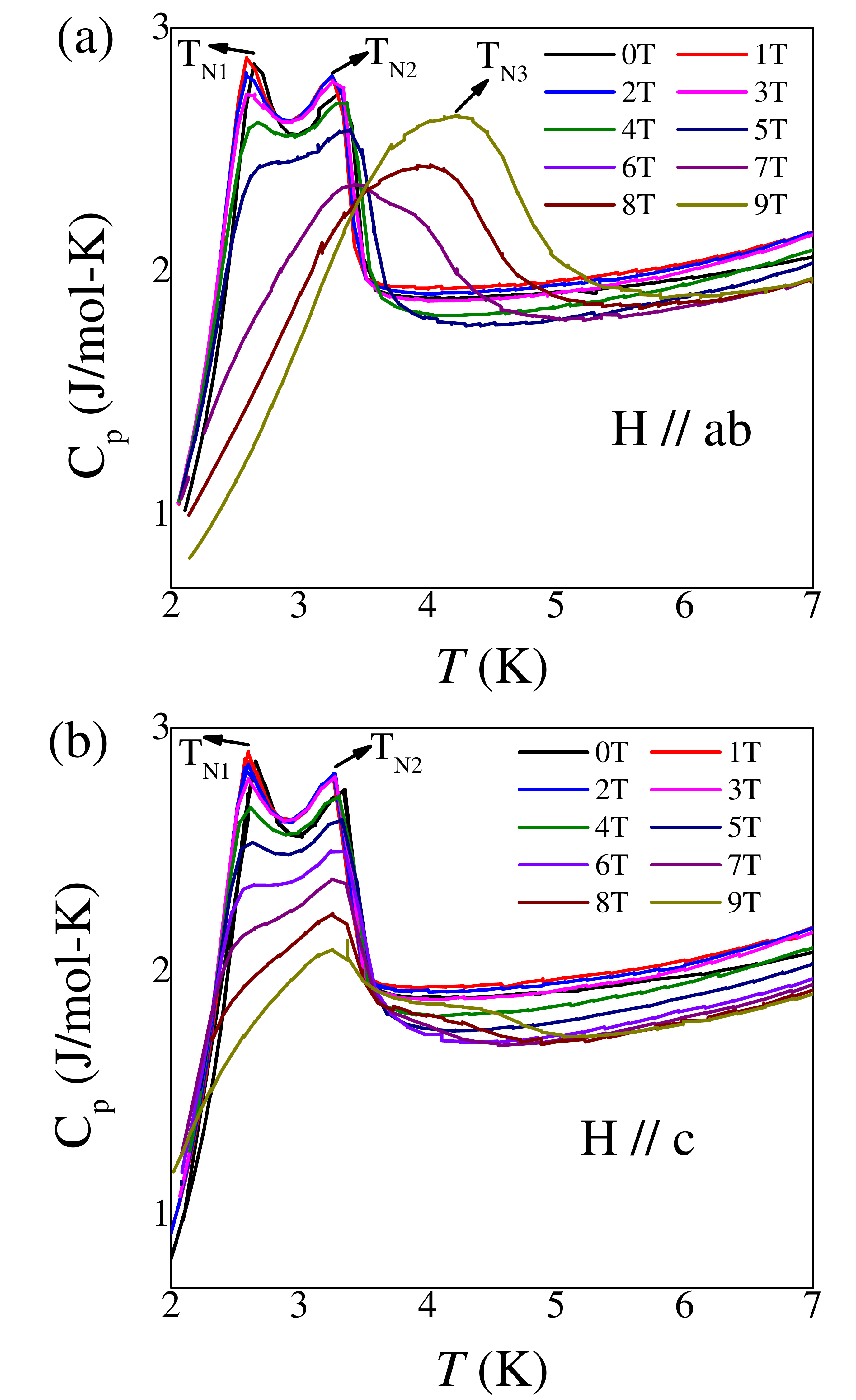}
	\end{center}
	\par
	\caption{\label{Fig:7} (color shown online) The temperature dependence of specific heat measured with $H \parallel ab$-plane (a) and $H \parallel c$-axis (b).}
\end{figure}

Fig. 7 shows the temperature dependence of the the specific heat, $C_{\text{p}}$, measured at different fields. At zero field, $C_{\text{p}}$ shows two sharp peaks at $T_{\text{N1}}$ = 2.7 K and $T_{\text{N2}}$ = 3.3 K, respectively, which indicates a two-step magnetic ordering. The $T_{\text{N1}}$ value is consistent with the ones determined by the neutron diffraction and $\chi(T)$ data.  With increasing field $H \parallel ab$-plane, these two peaks at $T_{\text{N1}}$ and $T_{\text{N2}}$ become less pronounced while keep almost unchanged positions. The peak at $T_{\text{N1}}$ disappears around 6.5 T while the peak at $T_{\text{N2}}$ disappears around 8 T. Meanwhile, around 7 T, another broad peak start to appear, whose position is labeled as $T_{\text{N3}}$ (Fig. 7(a)). This peak becomes stronger and shifts to higher temperatures with increasing field.  For $H \parallel c$-axis (fig. 7(b)), the peak at $T_{\text{N1}}$ becomes weaker and is unrecognizable around 8 T with increasing field. Meanwhile, the peak at $T_{\text{N2}}$ doesn't change obviously. Apparently, the evolution of $C_{\text{p}}$ under field is quite different between the $H \parallel ab$-plane and $H \parallel c$-axis cases.

\subsection{D. Magnetic phase diagrams}

\begin{figure}[tp]
	\linespread{1}
	\par
	\begin{center}
		\includegraphics[width= 3.0 in]{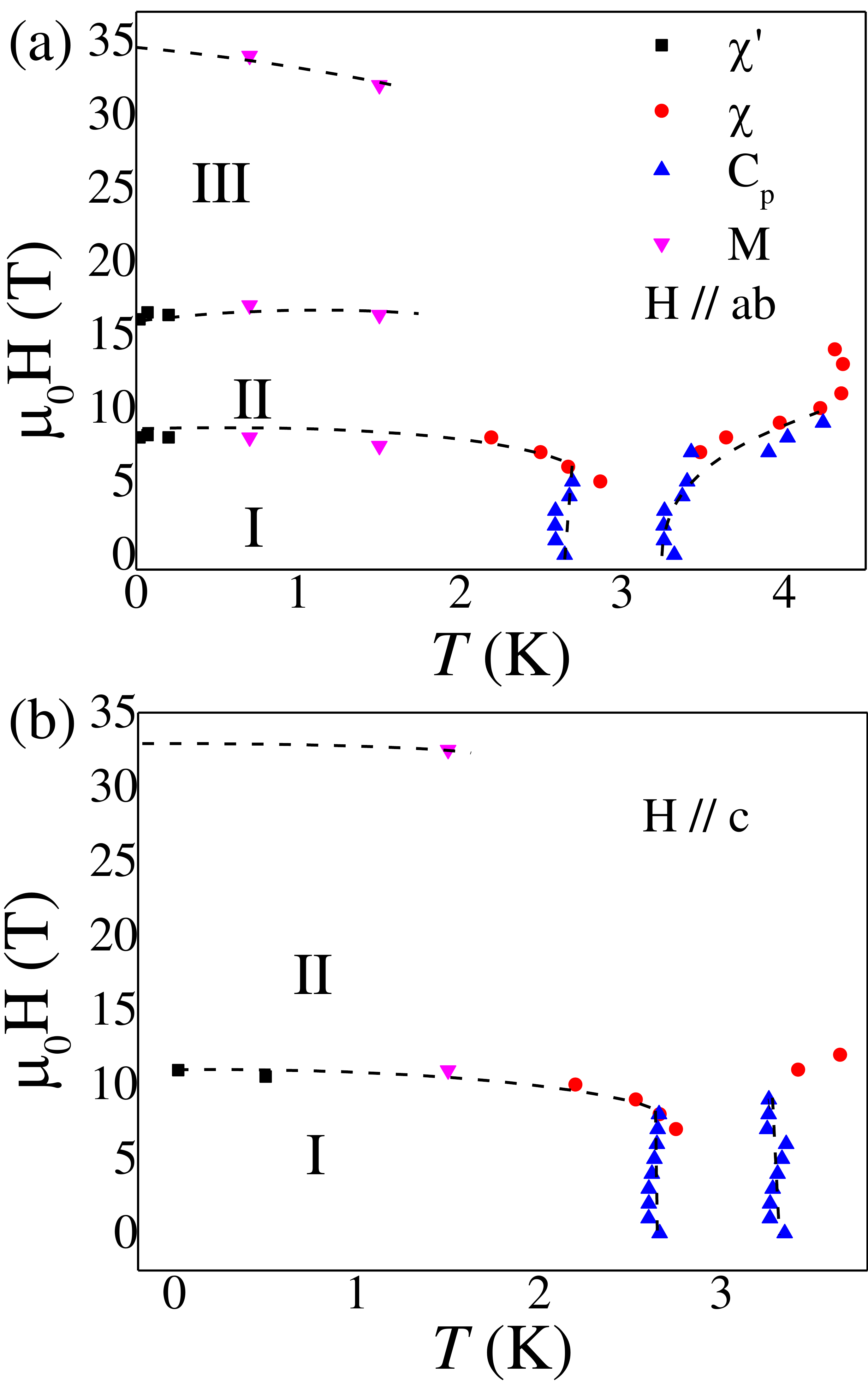}
	\end{center}
	\par
	\caption{\label{Fig:8} (color shown online) The magnetic phase diagram for $H \parallel ab$-plane (a) and $H \parallel c$-axis (b).}
\end{figure}

Magnetic phase diagrams for $H \parallel ab$-plane and $H \parallel c$-axis were constructed by using the critical temperatures and fields obtained above, as shown in Fig. 8. Besides the high temperature paramagnetic phase and high field fully polarized phase, the $H \parallel ab$-plane phase diagram includes three phases while the $H \parallel c$-axis one has two phases.

\subsection{E. Inelastic neutron scattering spectra}
	\begin{figure*}[tp]
	\centering
     {
		\includegraphics[width=6.5 in]{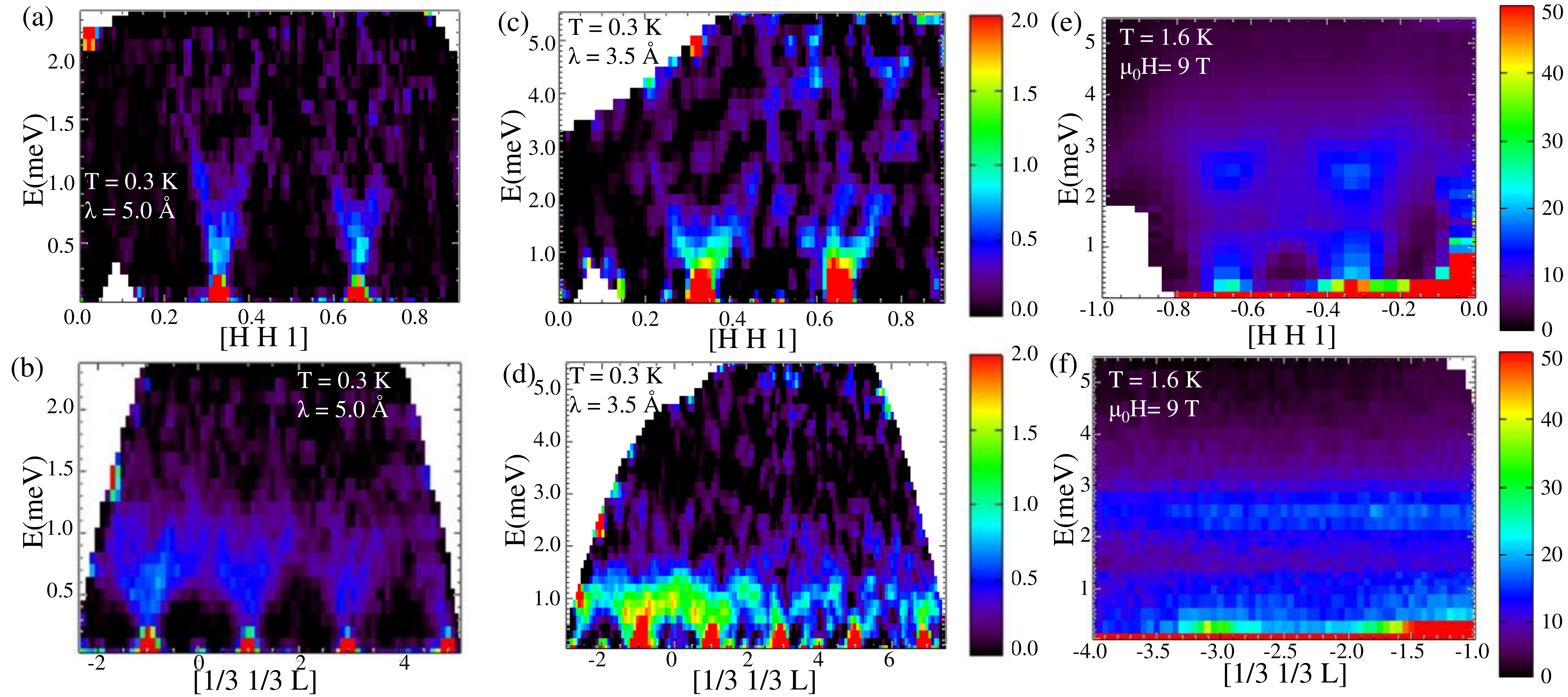}  
		\caption{\label{Fig:9}  The zero field INS spectra along the [H H 1] (a) and [1/3 1/3 L] (b) directions, which were measured at 300 mK using the wavelength $\lambda$ = 5$\, $\AA $\,$; For (c, d), the zero field INS spectra were measured at 300 mK using the wavelength $\lambda$ = 3.5$\, $\AA $\,$.  The INS spectra measured at 1.6 K with $H \parallel ab$-plane and $\mu_0H$ = 9 T along [H H 1] (e) and [1/3 1/3 L] (f) directions.}
	}
\end{figure*}

	\begin{figure*}[tp]
	\centering
     {
		\includegraphics[width=7 in]{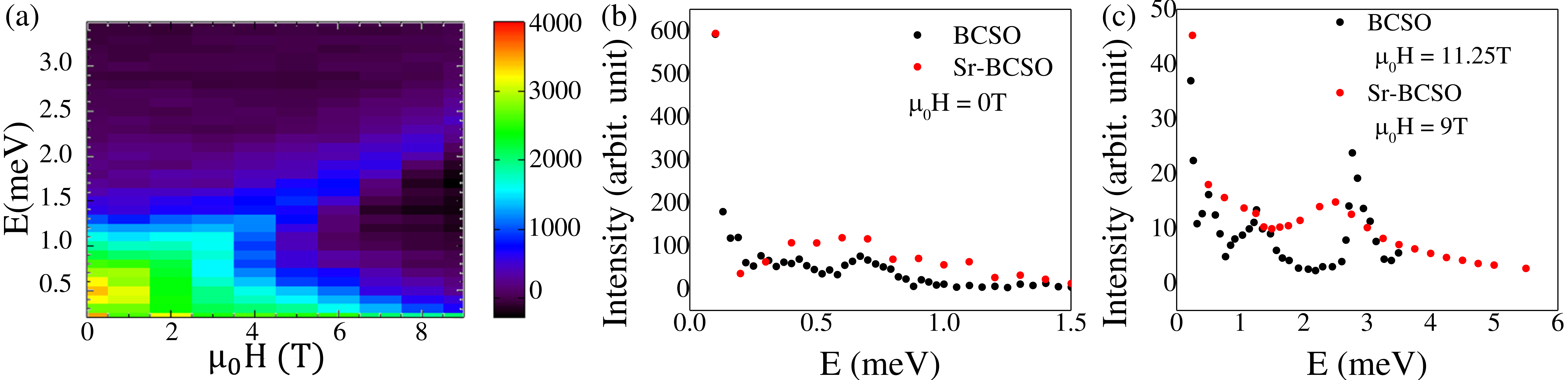}  
		\caption{\label{Fig:10}  (a) The energy cut at [1/3 1/3 1] as the function of magnetic field with $H \parallel ab$-plane. Comparison of the energy cut between Sr-BCSO and BCSO at the [1/3 1/3 1] position at zero field (b) and applied field (c), which is 9 T for Sr-BCSO and 11.5 T for BCSO.}
	}
\end{figure*}

The INS spectra of Sr-BCSO were measured at zero field and 9 T in the $ab$ plane, which allows us to study the spin dynamics in phase I and II in Fig. 8 (a). Fig. 9(a) and (b) show the INS spectra dispersion up to $~$ 2.5 meV along the [H H 1] and [1/3 1/3 L] directions at zero field, respectively. While the intralayer dispersion is sharp and the interlayer dispersion is quite broad, both of them only show one gapped mode. The energy cut at the zone center of [1/3 1/3 1] plotted in Fig. 10(b) shows no intensity near zero energy but a broad peak around 0.5 meV, which further confirms that there is only one gapped mode with the gap around 0.5 meV for the INS spectra. Another noteworthy feature is that there is no intensity at the M point for the intralayer dispersion. Fig 9 (c) and (d) show the INS spectra up to 5 meV along the two directions, both of them show no obvious intensity at energy above 1.5 meV.

The spectra under 9 T are shown in Fig. 9(e) and (f) for the two directions, respectively. It is obvious that there are two modes of broad dispersion along the H direction but two flat modes along the L direction. The energy cut at [1/3 1/3 1] plotted in Fig. 10(c) further confirms that one mode is gapless and another mode is gapped with a gap around 2.5 meV. The more detailed field scan of the cut at [1/3 1/3 1] is shown in Fig. 10(a). It shows that with increase field, the gapped mode at zero field splits to two modes, of which one's gap decreases to zero with field above 6 T and the other's gap increases. 

\subsection{IV DISCUSSIONS}

To learn the non-magnetic ion site disorder effects, we compare the observed magnetic properties of Sr-BCSO to those of the parent compound BCSO on several aspects.

(i) \emph{Anisotropy} The anomaly at $T_{\text{N1}}$ for $\chi(T)$ measured at low fields such as 5 T and the magnetization plateau feature are only observed for $H \parallel ab$-plane in Sr-BCSO. Both features are similar to those of BCSO, which indicates that the Sr-BCSO has the same anisotropy, the easy-plane anisotropy, as BCSO.

(ii) \emph{Magnetic ground state} The magnetic ground state for Sr-BCSO is a two-step transition with a weak ferromagnetic moment, or a possible canted 120 degree spin structure, as evidenced by the appearance of the [1/3 1/3 0] and [1/3 1/3 2] magnetic peaks at zero field. This is different from that of BCSO, which is a one step transition with a 120 degree spin structure showing no ferromagnetic moment\cite{sera2016s,zhou2012successive,ma2016static}. In BCSO, the [1/3 1/3 0] and [1/3 1/3 2] magnetic peaks only appear under applied fields. The room temperature single crystal diffraction measurement confirms that Sr-BCSO is isostructural to BCSO with an equilateral Co$^{2+}$ triangular layer. The low temperature single crystal neutron diffraction measurements further show no evidence of structural distortion below the magnetic ordering temperatures for Sr-BCSO. Moreover, as argued in (i), both samples have the same easy-plane anisotropy. Therefore, it is surprising to see that the Sr-doping can lead to such a different magnetic ground state in Sr-BCSO. 

One possible scenario is that the Sr doping can cause local structural distortion of the CoO$_6$ octahedra and therefore introduce a spatial anisotropy into the system. This spatial anisotropy, on top of the easy-plane anisotropy, can lead to a two-step transition with canted spins. For example, Ba$_3$NiNb$_2$O$_9$ has an equilateral triangular lattice for spin-1 Ni$^{2+}$ ions and orders at 4.9 K with a 120 degree spin structure in the $ab$ plane\cite{lu2018lattice}. By replacing Ba$^{2+}$ ion with Ca$^{2+}$ ion with smaller ionic size, a structural distortion is introduced into the system and therefore Ca$_3$NiNb$_2$O$_9$ now has a isosceles triangular lattice for Ni$^{2+}$ ions. One consequence is that Ca$_3$NiNb$_2$O$_9$ orders with a two-step transition at 4.2 K and 4.8K\cite{lu2018lattice}. Future studies on the local structure of Sr-BCSO is desirable to exam this possibility and better understand its magnetic ground state.

(iii) \emph{Ordering temperature and ordered moment} The ordering temperature of Sr-BCSO is around 3.3 K, which is 13\% reduction of the ordering temperature, 3.8 K, for BCSO. From the structural view, Sr-BCSO should have higher ordering temperature due to its shorter Co-Co intralayer distance since the Sr doping leads to a smaller lattice parameter $a$, which is 5.847 $\, $\AA $\,$. For BCSO, it is 5.856 $\, $\AA $\,$. Therefore, this lower ordering temperature could be related to the randomization of exchange interactions due to the Ba/Sr mixture. Meanwhile, the ordered moment for Sr-BCSO is around 1.24 $\mu_B$/Co, which is smaller than the 1.5 $\mu_B$/Co of BCSO. For BCSO, we also performed single crystal neutron diffraction measurements and obtained this ordered moment through refinement by using a 120 degree spin structure (the data is not shown here). In fact, both of these ordered moments are smaller than the saturation moment around 1.8 $\sim$ 2 $\mu_B$/Co. In a spin-1/2 triangular lattice antiferromagnet, the reduced ordered moment could be due to the quantum spin fluctuations. Therefore, the even smaller ordered moment of Sr-BCSO indicates that the Sr-doping enhances such kind of spin fluctuations. This enhancement could be another factor for the reduced ordering temperature.

(iv) \emph{Field induced spin state transitions} For $H \parallel ab$-plane, it is noted that the magnetization at $H_{\text{a1}}$ and $H_{\text{a2}}$ for Sr-BCSO  is 0.5 $\mu_B$ and 1.1 $\mu_B$, respectively, which is around 1/3, and $\sqrt3/3$ of the saturation value (1.8 $\mu_B$, Fig. 5(a)). Meanwhile, for BCSO, with increasing field along the $ab$ plane, its 120 degree spin structure at zero field is followed by a canted 120 degree spin structure; the UUD phase; a coplanar 2:1 canted phase (V phase) with one spin in the 120 degree spin structure rotated to be parallel with another spin, which gives $\sqrt3/3$ $M_{\text{s}}$; and another coplanar phase (V' phase) before entering the fully polarized state\cite{shirata2012experimental, susuki2013magnetization, sera2016s,zhou2012successive,gekht1997JETP,chen2013groundstates, starykh2014nearsaturation,yamamoto2014quantum,koutroulakis2015quantum, yamamoto2015microscopic,liu2019microscopic}. Therefore, here we tend to assign the phases I, II, and III for $H \parallel ab$-plane of Sr-BCSO as the canted 120 degree spin structure, the UUD phase, and the V phase. For Sr-BCSO, no V' phase was observed. It is obvious that the magnetization plateau feature in Sr-BCSO is much weaker than that of BCSO, which could be due to the Sr-doping disorder effect. In another triangular lattice antiferromagnet Rb$_{1-x}$K$_x$Fe(MoO$_4$)$_2$, the disorder introduced by the K-doping also weakens the magnetization plateau feature related to the UUD phase\cite{smirnov2017order}.

As learned from BCSO, while for $H \parallel c$-axis, its 120 degree spin structure will be followed by an umbrella spin structure and a V phase. However, for Sr-BCSO, the magnetic ground state is the canted 120 degree spin structure, which also should be the spin structure for phase I in its $H \parallel c$-axis phase diagram. Therefore, it is difficult to make the analogy between the $H \parallel c$-axis phase diagrams for Sr-BCSO and BCSO. The nature of the phase II for Sr-BCSO needs future studies to clarify.

(v) \emph{Spin dynamics} The observed INS spectra for Sr-BCSO have several differences from those for BCSO. (a) At zero field, only one gapped mode for Sr-BCSO but two modes (one gapless mode and one gapped mode with gap around 6 $\sim$ 7 meV) for BCSO\cite{ito2017structure,verresen2019avoided,ma2016static,macdougal2020avoided}; (b) Under field in the UUD phase, two modes for Sr-BCSO but three modes for BCSO\cite{kamiya2018nature}. The comparison of the energy cut at the [1/3 1/3 1] position (Fig. 10 (b, c)) further highlights these two differences. It is noticed that the two modes at low energy in the UUD phase for BCSO (the data measured at 11.25 T) may merge into each other in the UUD phase for Sr-BCSO (the data measured 9 T), as shown in Fig. 10(c); (c) No intensity at M point for Sr-BCSO but a flat mode and a rotonlike minimum at M point for BCSO\cite{ito2017structure,verresen2019avoided,ma2016static,macdougal2020avoided}; (d) No intensity above 1.5 meV for Sr-BCSO but a columnar continuum extending to at least 6 meV for BCSO\cite{ito2017structure,verresen2019avoided,ma2016static,macdougal2020avoided}. 

For BCSO, the modes at zero field correspond to the gapless Goldstone mode associated with rotation of the spins
in the $ab$ plane and an out-of-plane mode that is gapped in the presence of an easy-plane anisotropy, respectively. The rotonlike minimum at M point has been related to a magnon decay\cite{ma2016static}, but is under debate\cite{macdougal2020avoided}. The origin for the continuum at high energy is not clear so far. As we proposed above, the Sr-doping possibly introduces local spatial anisotropy to account for its two-step magnetic ordering. This spatial anisotropy will break the rotational symmetry of the spins in the $ab$ plane. Whether such kind of effect will lead to the disappearance of the gapless mode, or furthermore, the disappearance of the intensity at M point and the continuum, needs future theoretical studies to clarify.  

In summary, the Sr-doping on the non-magnetic Ba$^{2+}$ ion sites in BCSO affects its magnetic properties profoundly, including the reduction of the ordering temperature and suppression of the magnetization plateau through the randomization of exchange interactions, the reduction of the ordered moment by enhancing quantum spin fluctuations, and the modification of the magnetic ground states, phase diagrams, and spin dynamics by possibly introducing extra spatial anisotropy through local structural distortion. This study serves as a good example for demonstrating that disorder on non-magnetic ion sites play a complex role on quantum magnetism beyond the expected disruption of the exchange interactions. In fact, such kind of disorder can be treated as an perturbation to efficiently tune the magnetic properties of quantum magnets.

\begin{acknowledgments}
Q. H. and H.D.Z. thank the support from NSF-DMR through Award DMR-2003117. A portion of this work was performed at the NHMFL, which is supported by National Science Foundation Cooperative Agreement No. DMR-1157490, the U.S. Department of Energy, and the State of Florida. E.F. and H.B.C. acknowledge the support of U.S. DOE BES Early Career Award No. KC0402020 under Contract No. DE-AC05-00OR22725. C.R.W. thanks the support from the National Science and Engineering Research Council of Canada (NSERC), and the Canadian Foundation for Innovation (CFI),and the Canada Research Chair programme (Tier II). C.M. thanks the support from NSERC. W.X. was supported by Beckman Young Investigator Award. This research used resources at the High Flux Isotope Reactor, a DOE Office of Science User Facility operated by the Oak Ridge National Laboratory. Access to MACS was provided by the Center for High Resolution Neutron Scattering, a partnership between the National Institute of Standards and Technology and the National Science Foundation under Agreement No. DMR-1508249. J.M. thanks the National Science Foundation of China (No. 11774223 and U2032213 ). The authors thank Martin Mourigal and his research group for providing help in susceptibility measurements. The identification of any commercial product or trade name does not imply endorsement or recommendation by the National Institute of Standards and Technology. 
\end{acknowledgments}

\bibliographystyle{apsrev}
\bibliography{Sr-BSCO}

\end{document}